\documentclass[twocolumn,aps,prl,10pt,amsmath,amssymb,nofootinbib,showpacs,superscriptaddress,floatfix]{revtex4-1}

\DeclareFontFamily{U}{rcjhbltx}{}
\DeclareFontShape{U}{rcjhbltx}{m}{n}{<->rcjhbltx}{}
\DeclareSymbolFont{hebrewletters}{U}{rcjhbltx}{m}{n}

\usepackage{graphicx}
\usepackage{color}
\usepackage[usenames,dvipsnames]{xcolor}
\usepackage[colorlinks=true,linkcolor=Red,citecolor=Green,linktoc=page]{hyperref}
\usepackage{multirow}
\usepackage{float}
\usepackage{flushend}
\usepackage{balance}
\usepackage[varg]{txfonts}
\usepackage{ulem}
\usepackage{fancyhdr}

\DeclareMathSymbol{\lamed}{\mathord}{hebrewletters}{108}

\begin{document}
\title{Combinatorial Quantum Gravity: Emergence of Geometric Space form Random Graphs}

\author{C.\,Kelly}

\affiliation{School of Engineering and Physical Sciences, Heriot-Watt University, Edinburgh EH14 4AS, UK}

\author{C.\,A.\,Trugenberger}

\affiliation{SwissScientific Technologies SA, rue du Rhone 59, CH-1204 Geneva, Switzerland}

\begin{abstract}
We review and extend the recently proposed model of combinatorial quantum gravity. Contrary to previous discrete approaches, this model is defined on (regular) random graphs and is driven by a purely combinatorial version of Ricci curvature, the Ollivier curvature, defined on generic metric spaces equipped with a Markov chain. It dispenses thus of notions such as simplicial complexes and Regge calculus and is ideally suited to extend quantum gravity to combinatorial structures which have a priori nothing to do with geometry. Indeed, our results show that geometry and general relativity emerge from random structures in a second-order phase transition due to the condensation of cycles on random graphs, a critical point that defines quantum gravity non-perturbatively according to asymptotic safety. In combinatorial quantum gravity the entropy area law emerges naturally as a consequence of infinite-dimensional critical behaviour on networks rather than on lattices. We propose thus that the entropy area law is a signature of the random graph nature of space-(time) on the smallest scales. 
	
\end{abstract}
\maketitle

\section{Introduction}
Ultraviolet (UV) fixed points of statistical mechanics models define renormalizable quantum field theories via the Wilson renormalization group \cite{zinn}. Here we review the evidence, first presented in \cite{tru}, that quantum gravity is defined by an UV fixed point for a graph model \cite{graphrev}. The asymptotic safety scenario \cite{safety} would thus be realized on networks, rather than traditional statistical mechanics models. 

In the traditional discrete approach to quantum gravity \cite{triang} a smooth background is assumed, which is then approximated by piecewise flat geometries on which curvature is computed by Regge calculus \cite{regge}. In \cite{tru}, instead, one of us first posited that the fundamental structures on Planckian scales are not smooth but, rather, graphs, on which even notions like Regge curvature are lost. Random graphs are generic metric spaces. When equipped with a Markov chain, like a probability measure, a purely combinatorial notion of Ricci curvature, first introduced by Ollivier \cite{olli1, olli2, olli3}, can be defined on such structures. This was used in \cite{tru} to define a model of purely combinatorial quantum gravity. This approach was subsequently pursued in \cite{loll}, where a modified version of the Ollivier curvature was introduced. 

Albeit in a simplified model, we will provide strong evidence that geometric space emerges from random graphs at a second-order phase transition driven by a combinatorial version of the Einstein-Hilbert action and corresponding to the condensation of elementary loops on the graphs. In the geometric phase the combinatorial Einstein-Hilbert action becomes its standard continuum version. One notable result is that, in this model, the entropy of quantum space automatically follows an area law. The posited critical point on graphs could thus be the origin of the famed area law for the entropy in quantum gravity. Note that a relation between geometry and the density of triangles (the so-called clustering coefficient \cite{graphrev}) has been also noted in the network literature \cite{krioukov}. In the simplified model considered here we will be dealing with squares but the general case can also be treated \cite{kelly}.

\section{The Ollivier curvature and the combinatorial Einstein-Hilbert action}
The continuum Ricci curvature is associated with two (infinitely) close points on a manifold, defining a tangent vector. 
It can be thought of as a measure of how much (infinitesimal) spheres around these points are, on average, closer (positive Ricci curvature) or more distant (negative Ricci curvature) than the two points at their centres.  Its combinatorial version, the Ollivier curvature \cite{olli1, olli2, olli3}, is a discrete version of the same measure. Consider two vertices $i$ and $j=i+e_{ij}$ separated by edge $e_{ij}$ on a graph. The Ollivier curvature compares the Wasserstein (or earth-mover) distance $W\left( \mu_i, \mu_j \right)$ between two uniform probability measures $\mu_{i,j}$ on the unit spheres around $i$ and $j$ to the distance $d(i,j)$ on the graph and is defined as
\begin{equation}
\kappa (i,j)= 1- {W\left( \mu_i, \mu_j \right) \over d(i,j)} \ .
\label{olli}
\end{equation}
The Wasserstein distance between two probability measures $\mu_1$ and $\mu_2$ on the graph is defined as
\begin{equation}
W\left( \mu_1, \mu_2 \right) = {\rm inf} \sum_{i,j} \xi(i,j)d(i,j) \ ,
\label{wasser}
\end{equation}
where the infimum has to be taken over all couplings (or transference plans) $\xi(i,j)$ i.e. over all plans on how to transport a unit mass distributed according to $\mu_1$ around $i$ to the same mass distributed according to $\mu_2$ around $j$ without losses, 
\begin{equation}
\sum_j \xi (i,j) = \mu_1(i) \ , \qquad 
\sum_i \xi (i,j) = \mu_2(j) \ .
\label{transplan}
\end{equation}

The Ollivier curvature is very intuitive but, in general not easy to compute and work with. Fortunately, it becomes much simpler for bipartite graphs \cite{olli3}, which have no odd cycles. Since the Ollivier curvature of an edge depends only on the triangles, squares and pentagons supported on that edge (a discrete form of locality) \cite{olli2} and there are no triangles and pentagons on bipartite graphs, one can use for all practical purposes the simpler version of the Ollivier curvature for bipartite regular graphs \cite{olli3}:
\begin{eqnarray} 
\kappa (i,j) &&= -{1\over d} \Big[ (2d-2) -|N_1(j)| 
\nonumber \\
&&+ \sum_{a} \left( |L_a(j)|-|U_a(i)| \right) \times {\bf 1}_{\left\{ |U_a(i)| < |L_a(j)|\right\} } \Big] _+\ ,
\label{riccibipartite}
\end{eqnarray}
where $N_1(i)$ denotes the set of neighbours of $i$ which are on a 4-cycle supported on $(ij)$, ${\bf 1}$ denotes the indicator function (1 if the corresponding condition is satisfied, 0 otherwise) and the undescript ``+" denotes $z_+ = Max(z,0)$ so that the Ollivier Ricci curvature for bipartite graphs is always zero or negative. The definition of $U$ and $L$ is as follows: suppose that $R(i,j)$ is the subgraph induced by $N_1(i) \cup N_1(j)$ and $R_1(i,j)$...$R_q(i,j)$are the connected components of $R(i,j)$. Then $U_a(i) = R_a(i,j) \cap N_1(i)$ and $L_a(j) = R_a(i,j) \cap N_1(j)$ for $a=1 \dots q$. 

Equipped with a combinatorial version of Ricci curvature we can now formulate a purely combinatorial version of the Einstein Hilbert action. First we define the combinatorial Ricci scalar as 
\begin{equation}
\kappa (i) = \sum_{j \sim i} \kappa (i,j) \ ,
\label{orscalar}
\end{equation}
where $\sim$ denotes the neighbours of $i$ on the graph. Then we obtain the combinatorial Einstein-Hilbert action simply as
\begin{equation}
S_{\rm EH} = - {1\over g} \sum_i \kappa(i) \ ,
\label{cehaction}
\end{equation}
where the sum runs over all the vertices of the graph and $g$ is a coupling constant with dimension 1/action. 
This expression for the Ollivier curvature still looks forbidding. However, as we will show in a moment, it will become extremely simple on the physical configuration space. 

\section{Configuration space and combinatorial quantum gravity}

To fully specify a combinatorial quantum gravity model we need, in addition to the action, a configuration space over which to sum in the partition function. The configuration space in our simplified model consists of all random bipartite graphs. There is, however a further restriction that must be taken into account. 
As mentioned above we will model the emergence of geometry by the condensation of elementary loops on the graphs. We have thus to remind ourselves that even the Bose condensation of point particles is not well defined in absence of interactions, because of the infinite compressibility of the condensate. In exactly the same way, the condensation of ``non-interacting" loops is unstable, since it leads to crumpling and disconnected graphs (baby universes). We will thus follow the same route as for point particles and introduce, as the simplest stability mechanism, a hard-core condition for elementary loops. However, while for point particles the meaning of a hard core condition is unequivocal, for loops we must define what exactly we have in mind. The definition we will use is that elementary squares on the graphs will be allowed to share one edge but not more. Note that two squares can share two edges without being identical: it is exactly these configurations that we exclude. 

When the hard-core condition is implemented, the Ollivier combinatorial curvature becomes really simple. Indeed, it is easy
to convince oneself that the second term in (\ref{riccibipartite}), involving the sum of connected components of a subgraph, only contributes for squares that share 2 edges. Indeed, for an isolated square $|N_1|=1$ for all vertices on the square. 
If an edge supports $N_s$ squares which do not share another edge, then $|N_1(i)| = |N_1(j)|= N_s$ and $|U_a(i)|=|L_a(j)|$ since all the vertices within $N_1(i)$ and $N_1(j)$ are disconnected because of the absence of triangles in a bipartite graph and all the vertices of $N_1(i)$ are disconnected from those in $N_1(j)$ since, by assumption, the edge does not support two different squares. The Ollivier combinatorial curvature reduces thus simply to 
\begin{equation}
\kappa (i,j) = -{1\over d} \big[ (2d-2) -N_s(ij) \big]_+ \ ,
\label{riccireduced}
\end{equation}
where $N_s(ij)$ is the total number of squares supported on edge $(ij)$. The full model of combinatorial quantum gravity can thus be specified as
\begin{equation}
{\cal Z} = \sum_{CF} {\rm exp} \left( - {1\over g\hbar} \sum_i \kappa(i) \right) \ ,
\label{combqg}
\end{equation} 
where CF denotes the configuration space of random regular bipartite graphs with squares satisfying the hard-core condition and $\kappa (i) $ given by (\ref{orscalar}) and (\ref{riccireduced}). 

\section{The classical limit and the mean field action} 
The classical limit $\hbar \to 0$ corresponds to the weak coupling limit of small $g$. In the quantum regime $\hbar g \gg 1$ the Boltzmann probability becomes uniform over all configuration space of random regular bipartite graphs.
Random regular bipartite graphs are locally tree-like, with very sparse short cycles governed by a Poisson distribution with mean $(2d-1)^l/$ for cycles of length $l$ on $2d$-regular graphs \cite{wormald}. The quantity $(2d-2)$, instead, is the number of squares supported on an edge in a ${\mathbb Z}^d$ lattice. The combinatorial Einstein-Hilbert action (\ref{cehaction}) thus favours the formation of squares on the graph until the amount corresponding to a ${\mathbb Z}^d$ lattice is reached, after which it vanishes. The number of squares based on an edge, however, can be larger than $(2d-2)$ even for graphs with hard core squares, indeed it can reach up to $(2d-1)$. This is the maximum that the ``mean field" version of the action (\ref{riccireduced}) without subscript $+$ \cite{tru} would favour in the classical limit. It can be shown, however, that configurations attaining this maximum split into large quantities of disconnected baby universes while the classical limit of the exact action avoids this fate and approaches a regular  ${\mathbb Z}^d$ lattice \cite{kelly}. The hard core condition is thus sufficient to stabilize space. 

On a ${\mathbb Z}^d$ lattice the optimal coupling for the Wasserstein distance between the two vertices at the extremities of an  edge is the translation along the lattice links connecting the unit balls around the two vertices. As derived in \cite{olli1}, the Ollivier curvature then becomes the average of the sectional curvatures of the planes defined by the original edge and each of the links defining the unit ball around one of the vertices. If we assign a length $\ell$ to each link of the lattice and scale this as $\ell = \ell_0 N^{-1/d}$, with $\ell_0$ a renormalization constant related to the Planck length, we obtain the formal continuum limit 
\begin{equation}
{1\over \hbar g} \sum_i \kappa(i) \to {1\over 2(d+2) \ell_0^{d-2}} {N^{1-2/d}\over \hbar g} \int d{\rm Vol} \ R \ .
\label{ren}
\end{equation} 
Both sides do of course vanish since the ${\mathbb Z}^d$ lattice is Ollivier flat and the Euclidean space ${\mathbb R}^d$ it approximates is Ricci flat. However, this formal continuum limit shows, first, that the combinatorial Einstein-Hilbert action goes over into the continuum Einstein-Hilbert action and, secondly, that this requires a scaling of $g\hbar \sim N^{1-2/d}$. We will return to this all-important scaling below. For the moment let us retain that the the combinatorial quantum gravity model has the correct formal continuum limit. 

Having established that the hard-core condition is sufficient to stabilize space and obtain the correct (formal) continuum limit we can adopt the simpler mean field action expressed in terms of the total number of squares provided we explicitly exclude configurations with more than $(2d-2)$ squares per edge. To this end we compute
\begin{eqnarray}
\sum_i \kappa (i) &&= -(4d-4) N + {1\over d} \sum_i \sum_{e_i} N_s\left( e_i \right) 
\nonumber \\
&&= {-8\over d} \left[ {d(d-1)\over 2} N-N_s \right] \ ,
\label{totcursca}
\end{eqnarray}
which gives the final result
\begin{equation}
S_{EH}^{mf}=  {4d-4\over g} N   \big[ 1  -\zeta \big] \ ,
\label{meanfield}
\end{equation}
where $\zeta = 2 N_s/(d(d-1)N)$ ($0\le \zeta \le 1$) is the density of squares. 

\section{A continuous network phase transition and the entropy area law}
Let us consider the free energy of the model (divided by the ``temperature" $\hbar g$) 
\begin{equation}
F=  {4d-4\over \hbar g} N   \big[ 1  -\zeta \big] -  S\left( N \right) \ ,
\label{freeenergy}
\end{equation}
where $S(N)$ is the entropy of the graphs. In traditional statistical mechanics models the degrees of freedom, typically (but not necessarily) living on the vertices of a lattice, interact with a fixed number of their neighbours. As a consequence, both the energy and the entropy are extensive quantities, scaling like the volume $N$ (number of vertices of the lattice). Phase transitions, thus, show up when the external intensive parameter temperature $T$ (or coupling constant in case of quantum phase transitions) crosses a critical value $T_c$ where energy and entropy exactly compensate. This is not so in the statistical mechanics of networks \cite{newman}. On networks, interactions are represented by edges. Each vertex can thus interact with a number of other vertices that diverges in the limit $N\to \infty$: contrary to traditional statistical mechanics models on lattices, network are infinite-dimensional. Moreover, there is no a priori notion of locality on networks. On random graphs, e.g., there can be an edge between any two vertices. As explained in detail in \cite{graphrev}, the infinite-dimensionality of networks has the consequence that the phase structure of networks is determined by {\it critical functions} of $N$ rather than critical values, e.g.
\begin{eqnarray} 
{\rm phase \ 1} \ \ {\rm if} \ \ &&{\rm lim}_{N\to \infty} \left({T(N)\over T_c(N)}\right) = 0 \ ,
\nonumber \\
{\rm phase \ 2} \ \ {\rm if} \ \ &&{\rm lim}_{N\to \infty} \left({T(N)\over T_c(N)}\right) = \infty \ .
\label{phasesnet}
\end{eqnarray}
When the temperature (or the coupling in the quantum case) is chosen to scale exactly as the critical function,
\begin{equation}
{\rm lim}_{N\to \infty} \left({T(N)\over T_c(N)}\right) = t \ ,
\label{crfct}
\end{equation}
the phase transition appears typically as a traditional one in terms of the rescaled coupling $t$, one phase appearing for $t<t_c$ and the other for $t>t_c$ with the difference that energy, entropy and free energy need not be extensive quantities anymore but can have a different scaling in terms of the volume $N$ (see e.g. the mean field solution of the two-star model in \cite{newman}). 

This is exactly what seems to be realized in the present model of combinatorial quantum gravity. The formal continuum limit (\ref{ren}) is well-defined only if the coupling $\hbar g$ scales as $\hbar g \sim N^{1-2/d}$. And remarkably, exactly when this scaling is chosen, the order parameter $\zeta = 2N_s/(d(d-1)N)$ representing the density of squares collapses onto an $N$-independent function suggestive of a traditional second-order phase transition, as shown in Fig. 1 for the case $d=4$. 

\begin{figure}
\includegraphics[width=8cm]{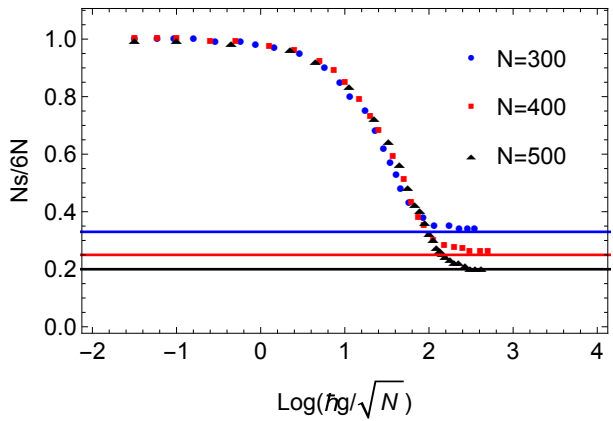}
\caption{\label{fig:Fig. 1} Monte Carlo simulation of the average number of squares for $d=4$ and $N=300$, $N=400$ and $N=500$ as a function of the rescaled coupling $\hbar g/\sqrt{N}$. Random regular graphs with sparse squares $N_s \sim {\rm Poisson\ } (600.25)$ and logarithmic distance scaling at large values of the coupling constant turn into $\mathbb{Z}^4$ lattices with the maximum number of squares $N_s=6N$ and power-law distances when gravitation becomes weak.The horizontal lines correspond to the expected number of squares for random regular graphs of the corresponding volume $N$.}
\end{figure}

Second order phase transitions are characterized by the divergence of the correlation length at the critical point. We can define a correlation length also on graphs. To this end we define the local order parameter $\zeta(i) = 2N_s(i)/(d(d-1))$ characterizing the density of squares at each vertex $i$ and we compute the correlation function 
\begin{equation}
C\left(d (ij) \right) = {<(\zeta (i) - \bar \zeta) (\zeta(j)-\bar \zeta) >\over \sigma (i)  \sigma  (j) } \ ,
\label{correl}
\end{equation} 
as a function of the graph distance $d(ij)$ between vertices. Here $\bar \zeta$ denotes the expectation value whereas $\sigma(i)$ is the standard deviation of $\zeta (i)$. Expressing, as usual, this correlation function as
\begin{equation}
C\left(d (ij) \right) = {\rm exp} \left( - {d(ij)\over \xi} \right) \ ,
\label{corle}
\end{equation}
defines the correlation length $\xi$. The correlation length in units of the graph diameter, averaged over different graph sizes, is plotted as a function of the rescaled coupling $ G= \hbar g$ in Fig. 2  for the case $d=3$.

\begin{figure}
\includegraphics[width=8cm]{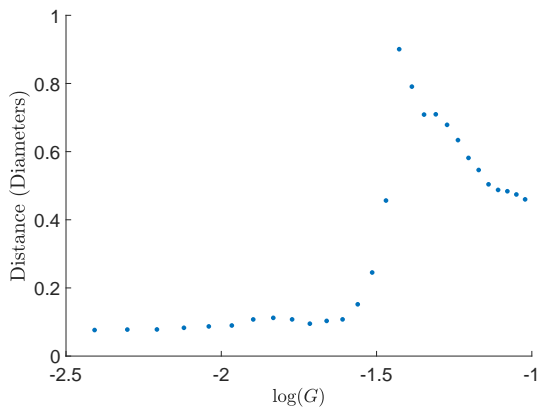}
\caption{\label{fig:Fig. 1} The correlation length in units of the graph diameter, averaged over different graph sizes as a function of the rescaled coupling $ G= \hbar g$ in Fig. 2  for the case $d=3$. }
\end{figure}

Remarkably, this plot indeed shows the typical behaviour of the correlation length at a second order phase transition, the divergence being of course cut-off by finite size effects. This second order phase transition, if confirmed by further studies, marks the emergence of geometric space, in the form of flat tori locally isomorphic to ${\mathbb Z}^d$ from random regular graphs as a consequence of the condensation of the shortest cycles, squares. This critical point would provide a non-perturbative definition of quantum gravity according to the asymptotic safety scenario \cite{safety}, in which the critical value 
$G_c \simeq 400 \ (d=4) $ (see Fig.1) of the rescaled coupling $\hbar g/N^{1-2/d}$ defines Newton's gravitational constant via the relation $G_{\rm Newton} = (3/4\pi) G_c \ell_0^2/\hbar$. 

A further important consequence of this second-order phase transition is derived from the expression (\ref{freeenergy}) of the free energy. As explained above, the scaling $\hbar g \sim N^{1-2/d}$ is exactly the critical function of the graph model. The existence of a second-order phase transition with this scaling would imply the existence of a balance critical value between energy and entropy, which would further imply that, with this scaling, energy and entropy have themselves the same scaling behaviour with the volume $N$. This immediately leads to the scaling behaviour $S(N) \sim N^{2/d}$ for the entropy of the graphs, in both phases of course. Since $N$ represents the volume, this scaling law is a combinatorial version of the entropy area law in quantum gravity \cite{area}.

\end{document}